\documentclass[twocolumn,aps,prd,amsmath,amssymb,preprintnumbers]{revtex4}
\begin{document}


\title{Comment on ``Towards a quantum notion of covariance in spherically symmetric loop quantum gravity''}

\author{Martin Bojowald}
\affiliation{Institute for Gravitation and the Cosmos,
The Pennsylvania State University, 104 Davey Lab, University Park,
PA 16802, USA}

\maketitle

The recent paper \cite{LoopSchwarzCov} by Gambini, Olmedo and Pullin (GOP)
promises new constructions that may make it possible to achieve covariance in
spherically symmetric models of loop quantum gravity. This claim is contrary
to the discovery of several stubborn obstacles to covariance uncovered in the
same models during the last few years
\cite{SphSymmCov,Disfig,NonCovPol,BlackHoleModels}. As the authors point out
in their abstract,
new progress is supposedly possible if one takes into account crucial features
of a certain partial Abelianization of the gauge transformations of
spherically symmetric gravity, published by the same authors in
\cite{LoopSchwarz}. In canonical form, these gauge transformations are given
by hypersurface deformations.  A closer inspection of the technical
calculations performed by GOP shows that spherically symmetric hypersurface
deformations are, in fact, violated in the construction.

Following \cite{SphSymm,SphSymmVol}, canonical spherically
symmetric gravity can conveniently be described by triad variables
$(E^x,E^{\varphi})$ which appear in the line element
\begin{equation}
  {\rm d}s^2= \frac{(E^{\varphi})^2}{E^x} {\rm d}x^2+ E^x ({\rm d}\vartheta^2+
  \sin^2\vartheta{\rm d}\varphi^2)
\end{equation}
in standard spherical coordinates. We assume a fixed orientation of the
triad such that $E^x>0$. Up to constant factors,
the momenta conjugate to $E^x$ and $E^{\varphi}$, respectively, are given by
extrinsic-curvature components $(K_x,K_{\varphi})$: We have
\begin{eqnarray}
  \{K_x(x),E^x(y)\}&=& 2G\delta(x,y)\\
  \{K_{\varphi}(x),E^{\varphi}(y)\}&=&G\delta(x,y)
\end{eqnarray}
with Newton's constant $G$. The coefficients on the right differ for the
radial direction $x$ and the angular directions $\vartheta$ and $\varphi$
because the latter amount to two fields that are equal by virtue of spherical
symmetry. It is possible to map to equal coefficients by redefining, say,
$K_x$ to $\tilde{K}_x=\frac{1}{2}K_x$, as implicitly done in
\cite{LoopSchwarzCov}, but this step is not essential for the arguments
presented there or here. In what follows, we will work with the original
extrinsic-curvature variables implied by symmetry reduction.

The basic fields are subject to the diffeomorphism constraint
\begin{equation}\label{D}
 D[M] = \frac{1}{G} \int{\rm d}x M(x) \left(-\frac{1}{2}(E^x)'K_x+K_{\varphi}'
   E^{\varphi}\right)\,,
\end{equation}
and the Hamiltonian constraint
\begin{widetext}
\begin{equation} \label{H}
 H[N]=\frac{-1}{2G}\int{\rm d}x N(x) \left(|E^x|^{-1/2}
   E^{\varphi}K_{\varphi}^2+
2 |E^x|^{1/2} K_{\varphi}K_x
+ |E^x|^{-1/2}(1-\Gamma_{\varphi}^2)E^{\varphi}+
2\Gamma_{\varphi}'|E^x|^{1/2}\right)
\end{equation}
where $\Gamma_{\varphi}=-(E^x)'/(2E^{\varphi})$ is a component of the spin
connection compatible with the densitized triad defined by $E^x$ and $E^{\varphi}$.

As is well known for canonical gravity in general, the Poisson bracket of two
Hamiltonian constraints have a rather complicated Poisson bracket with
structure functions. In spherically symmetric models, as found in
\cite{LoopSchwarz}, this structure can be simplified if one replaces $H[N]$
with the linear combination
\begin{eqnarray} \label{CHD}
  C[L]&=&H\left[\frac{(E^x)'}{E^{\varphi}} \int E^{\varphi}L{\rm
    d}x\right]-2D\left[K_{\varphi}\frac{\sqrt{E^x}}{E^{\varphi}} \int E^{\varphi}L{\rm
          d}x\right]\\
  &=& -\frac{1}{G}
\int{\rm d}x L(x) E^{\varphi}\left(\sqrt{|E^x|}
 \left(1+K_{\varphi}^2- \Gamma_{\varphi}^2\right)+{\rm const.}\right) \label{CL}
\end{eqnarray}
\end{widetext}
of the two original constraints. Here, $\int E^{\varphi}L{\rm d}x$ is a function of $x$ obtained
by integrating $E^{\varphi}L$ from some fixed initial point up to $x$. The new
constraints are Abelian:
\begin{equation} \label{CC}
  \{C[L_1],C[L_2]\}=0\,.
\end{equation}
The Abelian nature, and in particular the absence of structure functions, is
expected to simplify quantization or modification procedures applied to the
constraints. However, it also obscures the question of covariance because the
constrained system defined by $D[M]$ and $C[L]$ does not directly implement
hypersurface deformations.

We first turn to modifications of the constraints and their analysis given in
\cite{LoopSchwarzCov}, and then briefly discuss the role of hypersurface
deformations in this setting. (For a more detailed discussion of relevant
features of hypersurface deformations, see \cite{AbelianStructures}.)  GOP use
two different kinds of modifications, a generalized dependence of $C[L]$ on
$K_{\varphi}$ of the form
\begin{widetext}
\begin{equation} \label{CLf}
 C[L]= -\frac{1}{G}
\int{\rm d}x L(x) E^{\varphi}\left(\sqrt{|E^x|}
 \left(1+f_1(K_{\varphi})- \Gamma_{\varphi}^2\right)+{\rm const.}\right)
\end{equation}
\end{widetext}
with a non-polynomial function $f_1$, and a spatial discretization of
phase-space functions and their derivatives. (There are spatial derivatives in
(\ref{CL}) through $\Gamma_{\varphi}$, and in the diffeomorphism constraint.)
Because the authors use a certain combination of solutions to the constraints
and gauge-fixing conditions, it turns out that only the latter modification
survives in the final expressions for line elements that are supposed to be
invariant.

As part of the gauge-fixing procedure, GOP replace $\sqrt{f_1(K_{\varphi})}$
in expressions derived from (\ref{CLf}) with the {\em classical} functions of
$x$ that $K_{\varphi}$ happens to equal in two well-understood sample
space-time slicings.
(See the paragraph before equation (22) in \cite{LoopSchwarzCov}.)
Through the gauge fixing, the procedure therefore
replaces $f_1(K_{\varphi})$ with $K_{\varphi}$ in a given slicing, and thereby
removes the original modification in an evaluation of the modified
constraints. It is then not surprising that certain classical transformations
between slicings can be extended to the modified case.  However, this
procedure cannot be correct because it would imply that the classical
constraint (\ref{CL}) and the modification (\ref{CLf}) always have equivalent
solutions, even though they generate inequivalent equations of motion and
inequivalent gauge transformations. GOP are not analyzing covariance of the
modified constraint (\ref{CLf}), as they claim, but rather of a discretized
constraint with the classical dependence on $K_{\varphi}$.

The gauge fixing is problematic for various other reasons as well. In
particular, after solving the constraint $C[L]=0$ for $E^{\varphi}$ as a
function of $K_{\varphi}$ and $E^x$, the authors claim that this solution is
an observable even though it clearly does not Poisson commute with $C[L]$.
(See for instance ``the parameterized observable that defines
$E^{\varphi}_j$'' in the paragraph before equation (22) in their paper.)
It commutes only after the gauge fixing has been implemented, eliminating
$K_{\varphi}$ in favor of a spatial function, but this replacement does not
turn a phase-space function into an observable. (If this were the case, any
kinematical phase-space function could be turned into an observable by
replacing terms that do not commute with the constraints by gauge-fixing
functions.
A simple counting of degrees of freedom reveals the incorrectness of this
result because every first-class constraint removes two kinematical degrees of
freedom from the list of potential observables.)

GOP's constructions seem to be modeled on the practice of
deparameterization common for instance in quantum cosmology. In order to
deparameterize a Hamiltonian constraint $C$, one introduces a scalar degree of
freedom, $\phi$ with momentum $p_{\phi}$, which does not have mass or
self-interactions and is therefore described by a constant potential $V$. The
resulting scalar Hamiltonian, $H_{\phi}=\frac{1}{2}p_{\phi}^2/a^3+ a^3V$ (with
the cosmological scale factor $a$), is added to $C$. The new degree of freedom
$\phi$ makes it possible to derive Dirac observables for $a$ and its momentum
$p_a$, depending on $\phi$. Since $p_{\phi}$ is constant in time for a
$\phi$-independent $V$, $\phi$ is a monotonic function of the gauge parameter
(such as proper time) associated with the combined Hamiltonian constraint. The
scalar itself can therefore be considered a gauge parameter, and choosing a
specific value of $\phi$ is a form of gauge fixing. In a field theory, one may
use position-dependent gauge-fixing functions.

In this situation, it is correct to say that every kinematical degree of
freedom in the original setting (before the scalar contribution was added) can
be turned into a Dirac observable. The counting of degrees of freedom works
out because one adds a pair of degrees of freedom, $\phi$ and $p_{\phi}$,
which are then removed as independent observables after the constraint has
been solved and its gauge factored out. However, no additional degree of
freedom such as $\phi$ is added to the gravitational phase space in
\cite{LoopSchwarzCov}. Imposing the constraints then must remove some of the
kinematical degrees of freedom from the list of possible observables, in
contradiction with what has been claimed there.

The scalar model is also useful to illustrate why GOP misidentify
$E^{\varphi}$ as a Dirac observable. From the point of view of Dirac
observables, their $E^{\varphi}$ is analogous to the momentum $p_{\phi}$ while
$K_{\varphi}$ is analogous to $\phi$. GOP solve the Abelianized constraint for
$E^{\varphi}$ in order to obtain what they claim is a Dirac observable. In the
scalar model, one can similarly solve the combined constraint, $C+H_{\phi}=0$,
for $p_{\phi}=\pm \sqrt{-2a^3(C+a^3V)}$. This expression is a Dirac
observable, but only if $V$ is indeed constant and does not depend on
$\phi$. In \cite{LoopSchwarzCov}, by contrast, the expression obtained for
$E^{\varphi}$ by solving the constraint depends on $K_{\varphi}$. If we had
used a scalar-dependent potential $V(\phi)$ in the cosmological model, the
resulting $p_{\phi}$ would not have been a Dirac observable because it does
not have a vanishing Poisson bracket with the combined constraint. Moreover,
it would clearly be gauge dependent because $p_{\phi}$,
according to its Hamilton's equation generated by a Hamiltonian constraint
with $\phi$-dependent potential,
cannot be constant in the direction of
the gauge-parameter $\phi$. (For further mathematical subtleties related to
deparameterization of $\phi$-dependent constraints, see \cite{AlgebraicTime}.)

This analogy illustrates why the identification in \cite{LoopSchwarzCov} of
$E^{\varphi}$ as a Dirac observable (without further explanations) is
incorrect. In fact, later constructions in that paper,
such as the space-time metric components (22)-(26) derived from solutions for $E^{\varphi}$,
implicitly confirm that $E^{\varphi}$, which indeed determines a metric
component, should not be gauge invariant. It should be gauge or slicing
dependent if it is to contribute to a metric that results in an invariant line
element when combined with coordinate differentials. It turns out to be gauge
dependent in later parts of the paper, but it should be gauge dependent in a
specific way for an invariant line element to result after the constraints
have been modified. By first misidentifying $E^{\varphi}$ as a Dirac
observable and then refraining from considering its transformation properties
before the gauge is fixed, GOP relinquish control over covariance
properties. Moreover, it is not clear how the expressions used can be
interpreted as operators in a meaningful way: They cannot act on the physical
Hilbert space because they are not Dirac observables, and they cannot be
kinematical operators because they are gauge fixed.

Another misidentified observable is the function $E^x$;
see for instance equation (8) in \cite{LoopSchwarzCov} where an operator for
$E^x$ is set equal to what has been introduced as a Dirac observable in (7).
GOP arrive at this
formal claim by applying a procedure from loop quantum gravity
\cite{LoopRep,ALMMT} in which phase-space variables are discretized in space
and represented on spin-network states. In spherically symmetric models
\cite{SymmRed,SphSymm}, the networks are 1-dimensional with labels on vertices
(used as eigenvalues for an operator $\hat{E}^{\varphi}$) and on connecting
links (used as eigenvalues for an operator $\hat{E}^x$). The diffeomorphism
constraint is then implemented not as a quantization of the phase-space
function (\ref{D}) but rather in a finite version that shifts the vertices by
finite amounts according to a spatial diffeomorphism. The spin-network labels
are invariant under this transformation, and $E^x$ also commutes with the
Abelianized constraint $C[L]$ or its quantum version. In this construction,
$\hat{E}^x$ may therefore be considered a Dirac observable.

However, there is an unresolved contradiction with the fact that the classical
$E^x$, simply related to a specific metric component, is not
diffeomorphism invariant as a spatial function, let alone as a component of a
space-time object. In canonical gravity, it is not a Dirac observable and does
not commute with the diffeomorphism constraint. GOP acknowledge this tension
only implicitly when they use, later on in the paper, additional gauge-fixing
conditions for the eigenvalues $k$ of $\hat{E}^x$, even though this operator
is supposed to be a Dirac observable that should be gauge invariant. (See
``gauge fixing such that $x^2_j = \ell_{\rm Planck}^2 k_j$'' with the vertex
position $j$ and $\hat{E}^x$-eigenvalues $k$ just before equation (10) in
\cite{LoopSchwarzCov} and ``in the gauge $E^x_j = {\rm sig}(j)x^2_j$'' just
before equation (34) there.)

Even if consistent gauge-fixing procedures were applied, the claims would be
problematic because GOP construct only a limited set of transformations for a
line element, which does not suffice to indicate general covariance. It is not
at all clear that the {\em same} transformation between discretized spatial
slices can be used for all possible quantities that must be invariant in a
covariant theory. GOP mention that ``We have also studied the covariance of
several curvature scalars: the Ricci and the Kretschmann scalars, and the
scalar obtained by contracting the Weyl tensor with itself. We checked that in
the approximation where $x_j$ is treated as a continuous variable, which
allows to use derivatives instead of finite differences, these scalars do not
depend on the choice of the gauge function $F(x)$. This gives robustness to
our model regarding its covariance.''  However, far from implying robustness,
the sentence describes a trivial observation because,
in the absence of a detailed discussion of higher-order derivative terms
implied by discretization effects,
this limit removes the
only non-classical effect (discretization) used in the explicit calculations,
after an earlier modification of the $K_{\varphi}$-dependence has been
eliminated by an unclear gauge fixing.
In spite of the title, \cite{LoopSchwarzCov} has not shown a single step
toward covariance because its assumptions and specific constructions are based
on simplifications that eliminate any non-trivial contributions.

We now briefly return to the question of how partially Abelianized constraints
and hypersurface deformations may be related. As shown in more detail in
\cite{AbelianStructures}, the partial Abelianization of \cite{LoopSchwarz},
used in \cite{LoopSchwarzCov}, cannot give a full picture of covariance. It is
obtained only for a restricted version of hypersurface deformations,
implicitly defined by using a phase-space dependent combinations of lapse and
shift in (\ref{CHD}) while Abelianization is obtained only for phase-space
independent $L$ in $C[L]$: Since $\Gamma_{\varphi}$ in (\ref{CL}) contains a
spatial derivative of $E^x$ and $C[L]$ depends on $K_{\varphi}$, (\ref{CC}) no
longer holds if $L_1$ or $L_2$ depend on $K_x$, or if they depend on
$(E^{\varphi})'$ or higher spatial derivatives. Phase-space dependence would
also re-introduce structure functions in the bracket (\ref{CC}) as well as in
the Poisson bracket of $C[L]$ with the diffeomorphism constraint. The absence
of structure functions requires a restriction on the possible phase-space
dependence of smearing functions in the constraints.  Within this restricted
setting, the transformation from $(H[N],D[M])$ to $(D[M],C[L])$, which relies
on phase-space dependent smearing functions, is not invertible. Covariance can
be recovered only if constructions in the partially Abelianized model can be
extended to unrestricted hypersurface deformations, but this condition, as
shown in \cite{SphSymmCov}, is realized only if the two appearances of
$K_{\varphi}$ in (\ref{CHD}) and (\ref{CL}), one in the smearing function of
the diffeomorphism constraint and one in the explicit phase-space dependence
of $C[L]$, are modified in a coordinated way. GOP, however, modify only the
latter but keep the former unchanged.

\medskip

\noindent The author thanks Rodolfo Gambini,
Javier Olmedo and Jorge Pullin for sharing a draft of \cite{LoopSchwarzCov}.
This work was supported in part by NSF grant PHY-1912168.


\end{document}